# 10-nm silicon nanostructures for phase-change UV-readable optical data storage

**Johann Toudert\* and Rosalía Serna**

*Laser Processing Group, Instituto de Óptica, CSIC, Madrid, Spain*
*\*Corresponding author: johann.toudert@csic.es*

**Abstract.** Achieving state-of-the-art optical data storage requires raising device capacity well above commercial standards. This requires media structured at a much smaller scale and enabling readout at a shorter wavelength. Current CDs, DVDs and Blu-rays are read with visible light, and are based on metallic reflection gratings and phase-change recording layers structured at the few-hundred-nm scale. Herein, we introduce 10-nm structured silicon as a promising UV-readable data storage platform. Recording on it harnesses the amorphous-to-crystalline phase-change of silicon, the two phases presenting well-constrasted UV optical properties. Furthermore, the phase-change contrast is strongly enhanced in the Vacuum UV thanks to the distinct interband plasmon resonances of the amorphous and crystalline nanostructures, which have an epsilon-near-zero and surface plasmonic character, respectively. Silicon nanogratings with a 10 nm width and a 20 nm period resonate near the wavelength of 120 nm, at which phase-change induces a 600% maximum optical transmittance contrast. This paves the way toward UV-readable data storage platforms with a 10 to 100 times increased data density, which could be implemented by harnessing the well-established silicon nanotechnology.

**Keywords:** silicon; phase-change materials; optical data storage; interband plasmonics; ultraviolet

# 1. Introduction

Achieving state-of-the-art optical data storage requires raising device capacity well above commercial standards. This requires media structured at a much smaller scale and enabling readout at a shorter wavelength. Current CDs, DVDs and Blu-rays are read with visible light (red or blue), and are based on metallic reflection gratings and phase-change recording layers structured at the few-hundred-nm scale. In these technologies, the different phases of the recording material display well-contrasted optical properties. Thus, locally triggered phase-changes enable writing optically-contrasted areas, which can subsequently be read with a laser beam scanning the surface.

While next generation devices could be inspired by such technologies, they must rely on other phase-change materials structured at the few-nm to few-tens-of-nm scale, and presenting strong UV optical contrasts to enable readout in this spectral range. When reviewing this specs list, one unavoidably thinks about employing phase-change chalcogenide plasmonic materials [1], which fulfill the conditions mentioned above. If, in addition, one wishes the chosen material to be widely available, it surprisingly (?) comes that a key candidate might be silicon.

Silicon (Si) is a widely abundant element, which is the cornerstone of many of the technologies used in daily life, spanning across the fields of electronics, optoelectronics, photonics, and photovoltaics. Si *is* a phase-change material, which can exist either in its amorphous phase (a-Si) or in its room-condition stable diamond cubic crystalline phase (c-Si). The transition between these phases can be triggered with heat, light, or electricity. Very recently, it was shown that both a-Si [2] and c-Si [3-12] nanostructures display so-called interband plasmonic properties [13-16], which are tunable by design across the Vacuum UV (VUV) and Middle UV (Mid UV) regions (wavelength from 100 to 300 nm). However, information about the achievable phase-change-induced UV optical contrasts between a-Si and c-Si nanostructures is still not available.

Herein, we thus explore the UV optical response of 10-nm structured silicon nanogratings in the a-Si and c-Si phases. We show that a-Si and c-Si nanogratings with the same geometry and dimensions display distinct interband plasmon resonances in the VUV region, enabling phase-change-induced optical transmittance contrasts of up to 600%. This positions 10-nm structured silicon as a promising UV-readable data storage platform.

# 2. Results

We first consider the simplified structure shown in **Figure 1a**, which consists of a periodic nanograting alternating Si nanostructures and air. Light is incident along the direction perpendicular to the nanograting. The Si nanostructures are semi-infinite along the beam and infinite perpendicular to it. Their width is w = 10 nm, and the nanograting period is P = 20 nm. Si may be amorphous or crystalline, with all the Si nanostructures being in the same phase. The VUV-to-Mid UV spectra of the real and imaginary parts of the dielectric function ($\varepsilon = \varepsilon_1 + i\varepsilon_2$) of a-Si and c-Si are also shown in **Figure 1a**. Their real part is negative ($\varepsilon_1 < 0$) in almost the whole range, i.e., they are optically metallic, as required for plasmonic properties to be achieved by nanostructure design. Interestingly, a-Si displays less negative $\varepsilon_1$ values than c-Si, with a near-cancelation at the wavelength of 120 nm, where a-Si thus displays epsilon-near-zero properties. The value of $\varepsilon_2$ increases with the wavelength for both a-Si and c-Si, due to the onset of their interband transitions peaking in the 300-400 nm range. In the 120-200 nm range, a-Si and c-Si display similar $\varepsilon_2$ values.

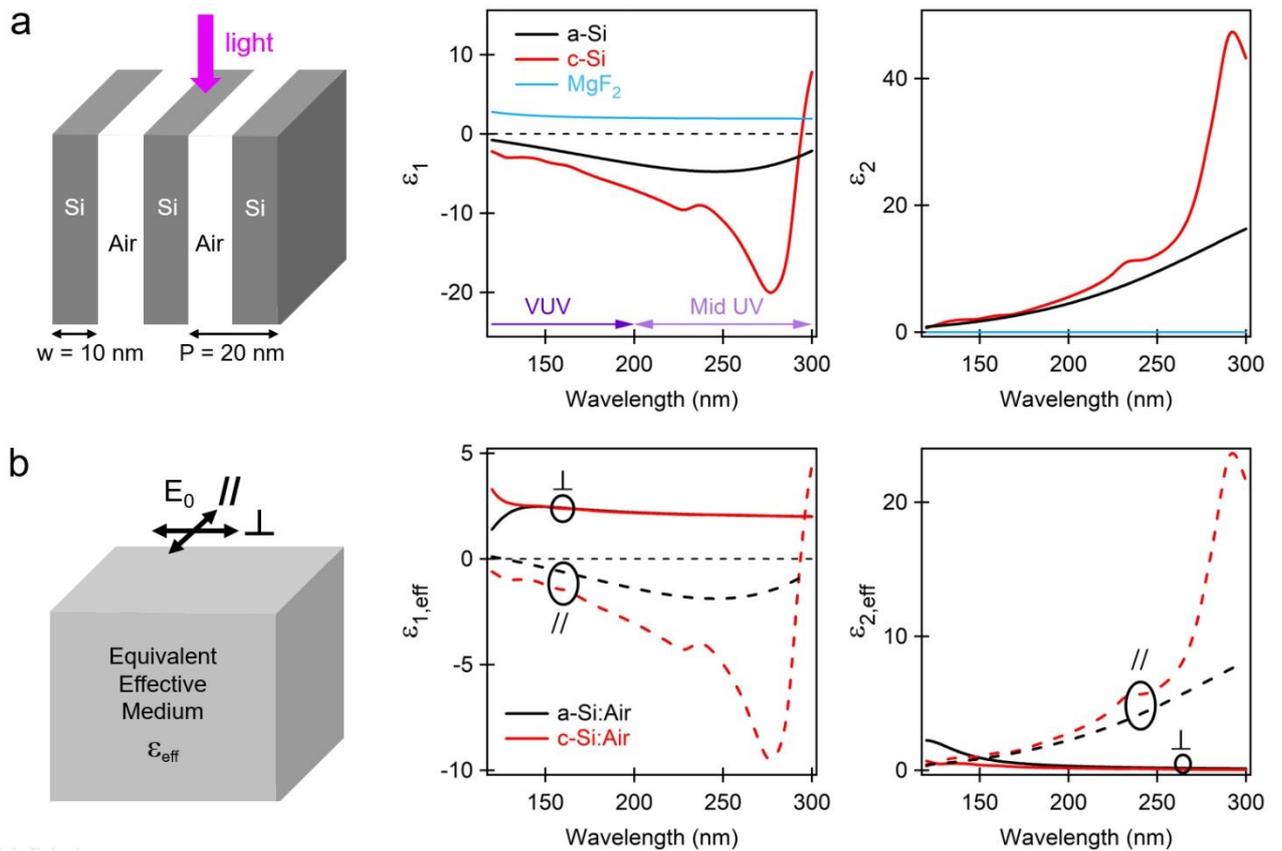

**Figure 1. Si nanograting: effective optical properties** (a) Nanograting alternating Si nanostructures and air. The nanostructure width is w = 10 nm, and the nanograting period is P = 20 nm. All the Si nanostructures are in the same a-Si or c-Si phase. VUV-to-mid UV spectra of the real and imaginary parts of the dielectric function ($\varepsilon = \varepsilon_1 + i\varepsilon_2$) of a-Si and c-Si, taken from the Palik database [17], and of MgF$_2$ adapted from [18]. (b) Equivalent effective medium. Spectra of its effective dielectric function $\varepsilon_{eff} = \varepsilon_{1,eff} + i\varepsilon_{2,eff}$ for the incident electric field E$_0$ linearly polarized in the direction parallel to the nanostructures (//) or perpendicular to them ($\perp$), for the Si nanostructures consisting of a-Si (a-Si:Air) or c-Si (c-Si:air).

As the nanostructure width is 10 nm, i.e., more than 10 times smaller than the considered wavelengths, the nanograting can as a first approximation be modeled as an equivalent effective medium, as shown in **Figure 1b**. Its effective dielectric function $\varepsilon_{eff}$ for the incident electric field $E_0$ polarized parallel to the Si nanostructures (//) or perpendicular to them ($\perp$) is derived from [19] as: $\varepsilon_{eff,//} = 0.5(\varepsilon_{Si} + \varepsilon_{Air})$ and $1/\varepsilon_{eff,\perp} = 0.5(1/\varepsilon_{Si} + 1/\varepsilon_{Air})$ **(Equation 1)**, with Si being a-Si or c-Si, and $\varepsilon_{Air} = 1$. The so-obtained spectra of the real and imaginary part of the effective dielectric functions are shown in **Figure 1b**. For the // polarization, $\varepsilon_{eff,//}$ is half the dielectric function of Si. The nanograting thus behaves as "dilute silicon", and remains optically metallic ($\varepsilon_{1,eff,//} < 0$) across the VUV and Mid UV. In contrast, for the $\perp$ polarization, both the real and imaginary parts of the effective dielectric function are positive ($\varepsilon_{1,eff,\perp} > 0$ and $\varepsilon_{2,eff,\perp} > 0$), so that the nanograting behaves as a lossy dielectric. Losses are stronger at VUV wavelengths below 150 nm, as a result of the plasmon resonance occuring in this region. Furthermore, higher $\varepsilon_{2,eff,\perp}$ values are observed in the a-Si case compared with the c-Si case, suggesting that the resonance leads to stronger losses in the a-Si nanograting compared with the c-Si one.

These differences in the $\varepsilon_{2,eff,\perp}$ values lead to an important a-Si to c-Si phase-change-induced transmittance contrast for the nanograting, provided it is few-tens-of-nm thick. To illustrate this fact, we consider a grating with a t = 40 nm thickness, standing on a UV-transparent 1 mm-thick $MgF_2$ substrate. The corresponding structure is shown in **Figure 2a**, together with its transfer matrix-calculated transmittance and reflectance spectra. These spectra reveal a strong anisotropy which results from the markedly different optically metallic and lossy dielectric effective properties in the // and $\perp$ polarized cases. Particularly relevant are the transmittance spectra in the $\perp$ polarized case, which confirm the presence of a VUV plasmon resonance, which is stronger for the a-Si nanograting than for the c-Si one, thus resulting in a near zero transmittance at the wavelength of 120 nm in the a-Si case, and in a 20% transmittance in the c-Si case. Accordingly, the a-Si to c-Si phase-change of the nanograting results in an important contrast of its VUV transmittance.

In order to ascertain these findings that rely on the validity of **Equation 1**, which is rigorously exact if the nanograting thickness is significantly larger than the Si nanostructure width (t >> w), we have performed full-wave FDTD simulations of the transmittance and reflectance spectra of the same structure. Technical details are provided in the **Methods Section**. The so-obtained spectra, shown in **Figure 2b**, are very similar to those obtained from effective medium simulations. This confirms the qualitative validity of the predicted trends, while providing quantitatively reliable results. The following discussions will thus be based on the data provided by FDTD simulations.

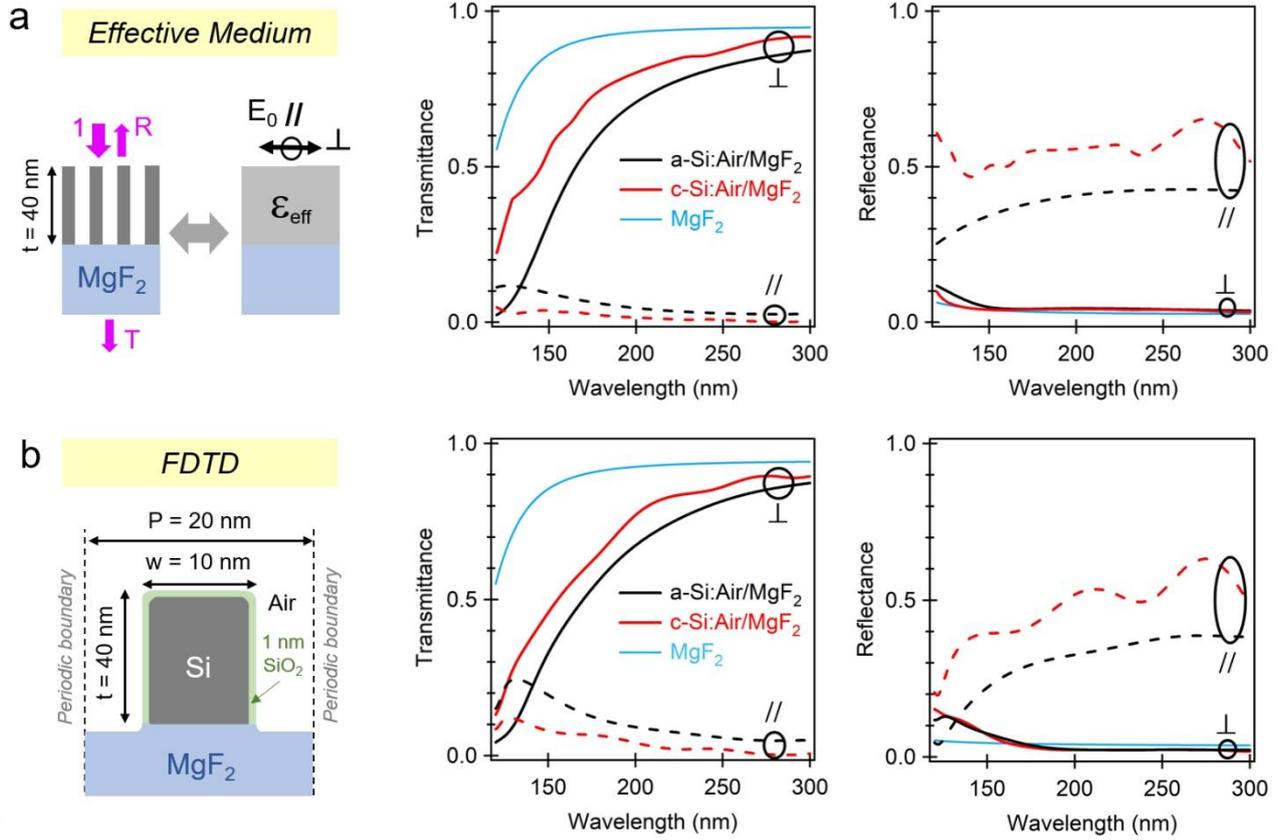

**Figure 2. Si nanograting on a MgF$_2$ substrate: transmittance and reflectance.** (a) Effective medium simulations. The nanograting is the same as in **Figure 1**, yet with a finite thickness t = 40 nm. The transmittance and reflectance are calculated for the // and ⊥ polarizations with the transfer matrix formalism, using the effective dielectric functions shown in **Figure 1b**, and the dielectric function of MgF$_2$ shown in **Figure 1a**. (b) FDTD simulations of the same structure: unit cell used and calculated transmittance and reflectance spectra. To account for periodicity, periodic boundary conditions were used. Refined details having a minor effect were added, such as rounded corners and a native 1 nm SiO$_2$ layer. The dielectric functions of a-Si, and c-Si were taken from [17], that of MgF$_2$ adapted from [18], and that of SiO$_2$ taken from [20], respectively. For the sake of readability, the aspect ratio of the unit cell is not represented at scale.

To address the physical origin of the a-Si to c-Si phase-change-induced transmittance contrast in the ⊥ polarized case, we have calculated by FDTD the electric field maps of the nanograting considered in **Figure 2b** (w = 10 nm, P = 20 nm, t = 40 nm), in the a-Si and c-Si cases, as a function of the wavelength λ. The obtained maps of the unit cell for λ = 120, 160, 200, 300 nm are represented in **Figure 3**. In the c-Si case, clear surface plasmonic features are observed, such as an enhanced field at the Si nanostructure's outer lateral surfaces for λ = 120 nm. The enhancement is greater near the top of the nanostructure, showing that the plasmon results from the coupling of the incident light impinging from the top of the cell. It then experiences an in-depth decay as a result of absorption losses in Si, into which the field penetrates. Upon increasing λ, the field does not penetrate anymore into Si, so that it becomes less attenuated and transmittance increases. In the a-Si case, a markedly different behaviour is observed. For λ = 120 nm, in contrast with c-Si, the field is not enhanced outside the Si nanostructure but inside it. This feature is characteristic of an epsilon-near-zero plasmonic behaviour [21-23]. It results in a stronger absorption loss in a-Si than in the case of c-Si, and therefore in a faster in-depth field decay and a lower transmittance. Then, upon increasing λ, the field enhancement pattern turns surface plasmonic, resulting in a reduced in-depth decay and a higher transmittance. The marked differences between the c-Si and a-Si cases can be explained qualitatively by the fact that the $\varepsilon_1$ of a-Si is less negative than that of c-Si in the VUV, thus allowing the epsilon-near-zero plasmonic behaviour and increased absorption loss in the a-Si case.

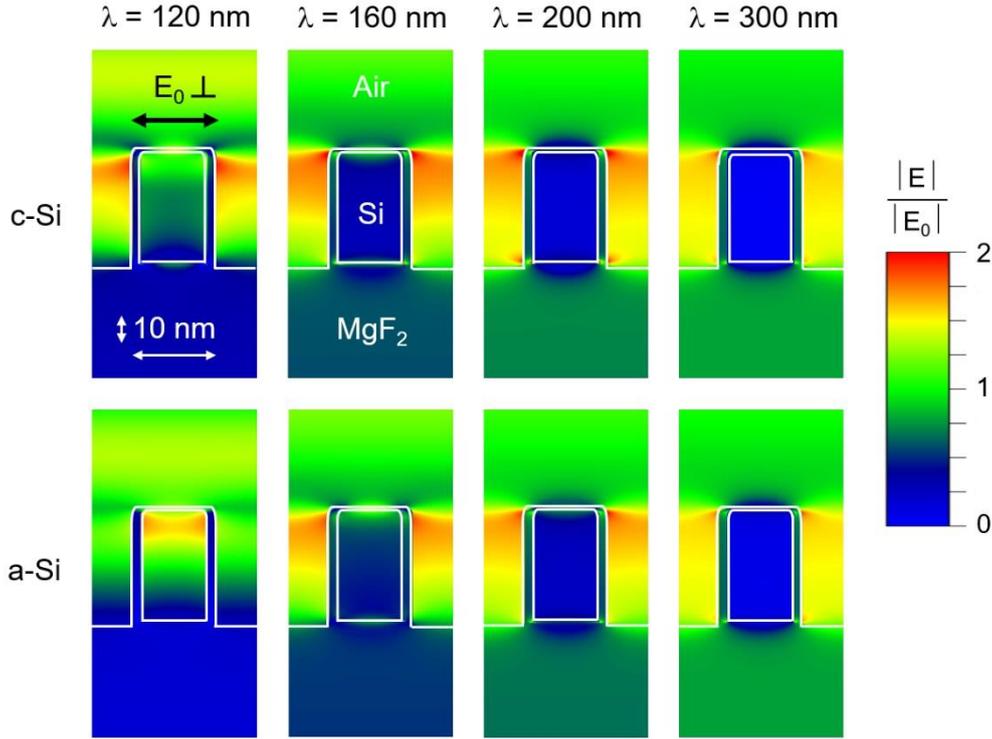

**Figure 3. ⊥ polarized electric field maps calculated by FDTD, as a function of the wavelength λ, for a c-Si and a a-Si nanograting on a MgF₂ substrate.** The unit cell and nanograting dimensions (w = 10 nm, P = 20 nm, t = 40 nm) are the same as in **Figure 2b**. The corresponding maps for the // polarized case are shown in the Supporting Information, **Figure S1**. For the sake of readability, the aspect ratio of the maps is not represented at scale. However, it is the same for all the maps to enable a direct comparison between them.

To further explore the interaction of ⊥ polarized light with the Si nanograting, we performed thorough FDTD simulations in which we varied its dimensions (thickness t, width w, period P = 2w). The transmittance spectra obtained in the c-Si and a-Si cases with varied thicknesses t = 10, 20, 40 and 60 nm (and w = 10 nm, P = 20 nm) are gathered in the Supporting Information, **Figure S2**. **Figure 4** shows the corresponding ⊥ polarized electric field maps at λ = 120 nm. Whatever the value of t, the maximum enhanced field is located at the outer lateral Si nanostructure surfaces in the c-Si case, and inside the Si nanostructure in the a-Si case, confirming their distinct plasmonic character (surface plasmonic and epsilon-near-zero plasmonic, respectively). Upon increasing t, the in-depth field decay becomes more clear, and it also becomes much stronger in the case of a-Si. Accordingly, the VUV transmittance of the nanograting decreases as t increases, as shown in the Supporting Information S2, and it becomes more markedly lower in the case of a-Si.

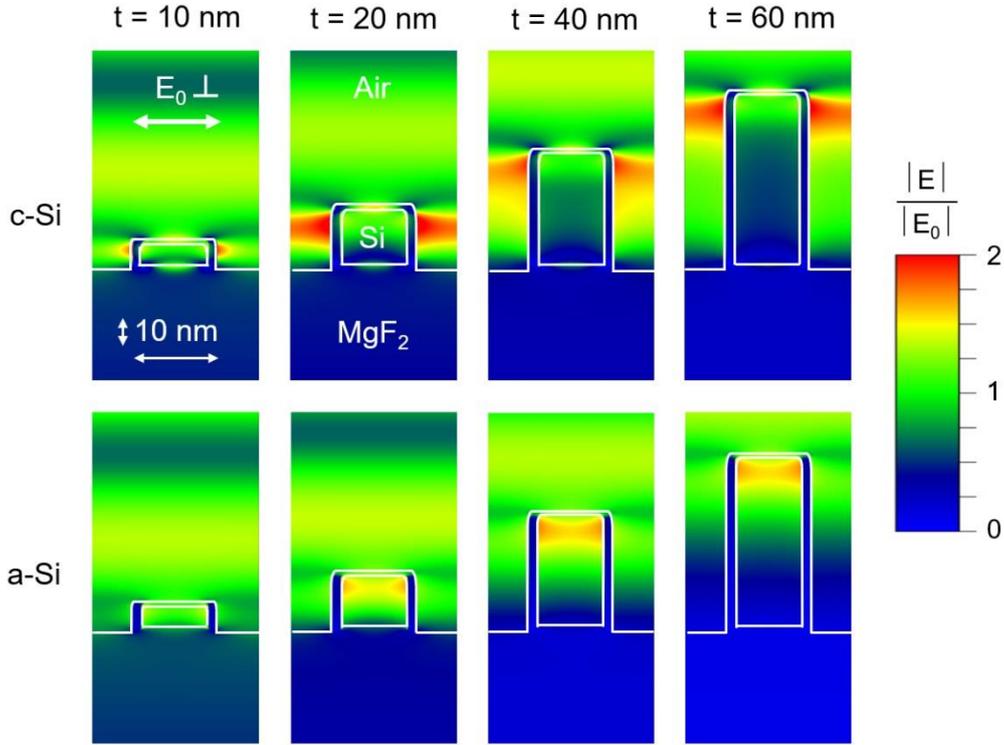

**Figure 4. ⊥ polarized electric field maps calculated by FDTD for a c-Si and a a-Si nanograting with thickness t = 10, 20, 40 and 60 nm on a MgF$_2$ substrate, at the wavelength λ = 120 nm.** The other nanograting dimensions are w = 10 nm and P = 20 nm. The corresponding transmittance spectra are shown in the Supporting Information, **Figure S2**. For the sake of readability, the aspect ratio of the maps is not at represented at scale. However, it is the same for all the maps to enable direct comparison between them.

The transmittance spectra obtained in the c-Si and a-Si cases for varied widths w = 5, 10 and 20 nm (and t = 40 nm, P = 2w) are gathered in the Supporting Information, **Figure S3**. **Figure 5** shows the corresponding ⊥ polarized electric field maps at λ = 120 nm. These maps prove that the field enhancement results from the cooperative effect of two plasmons localized at the opposite lateral surfaces of each Si nanostructure. These two plasmons are well separated for the nanostructures with w = 20 nm, and overlap upon decreasing w. This effect is particularly appealing in the a-Si case, where enhanced fields corresponding to the two epsilon-near-zero plasmons fully overlap inside Si for w = 20 nm. As a consequence of this effect, the absorption loss in Si markedly increases as w decreases, and thus the VUV nanograting transmittance decreases, as shown in the Supporting Information, **Figure S3**.

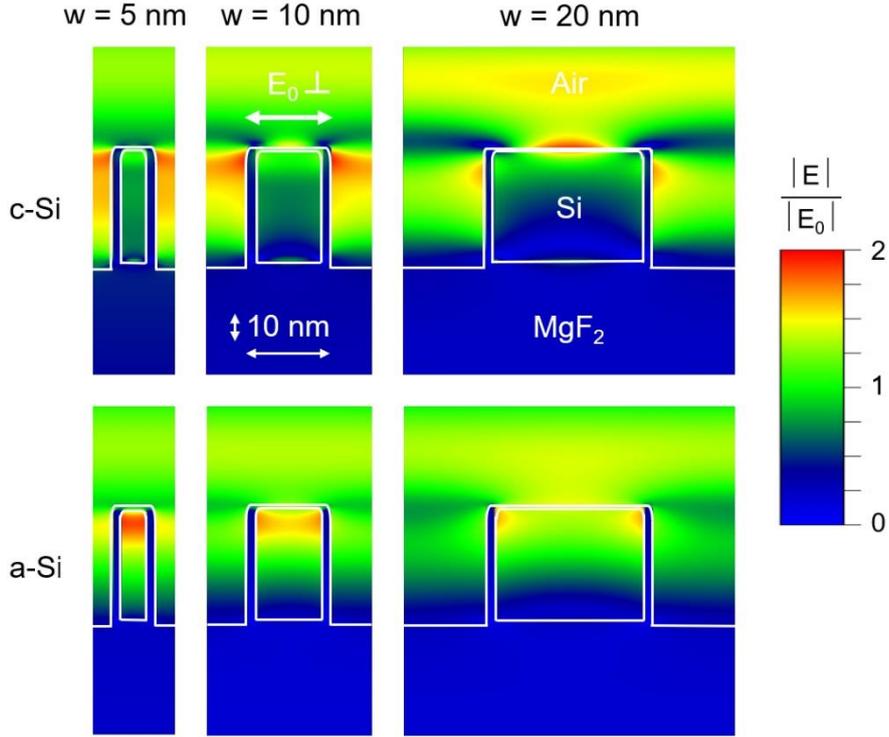

**Figure 5.** ⊥ polarized electric field maps calculated by FDTD, for a c-Si and a a-Si nanograting with Si nanostructure width w = 5, 10, 20 nm on a MgF$_2$ substrate, at the wavelength λ = 120 nm. The other nanograting dimensions are P = 2w and t = 40 nm. The corresponding transmittance spectra are shown in the Supporting Information, **Figure S3**. For the sake of readability, the aspect ratio of the maps is not at scale. However, it is the same for all the maps to enable direct comparison between them.

From the field maps shown in **Figure 4** and **Figure 5**, it is clear that the nanograting thickness t, Si nanostructure width w markedly influence the grating's interaction with light and its UV transmittance properties. To determine how these parameters affect the a-Si to c-Si phase-change-induced transmittance contrast between a-Si and c-Si nanostructures, we have calculated its spectrum from the data gathered in the Supporting Information, **Figures S2 and S3**, using the relation: Contrast(λ) = 100[$T_{cSi}$(λ) − $T_{aSi}$(λ)]/$T_{aSi}$(λ) (Equation 2), where $T_{cSi}$ and $T_{aSi}$ are the transmittance of the nanograting in the c-Si and a-Si case, for a given set of t, w, P values. The obtained spectra are plotted in **Figure 6**, where it is seen that whatever the t and w value, a maximum contrast is achieved resonantly near the wavelength of 120 nm. The contrast therefore results from the markedly different character of the plasmon resonances of the c-Si nanograting (surface plasmon) and a-Si nanograting (epsilon-near-zero plasmon). The maximum value increases when t increases and w decreases, with a maximum value of 600% being obtained for t = 60 nm, w = 10 nm, and P = 20 nm.

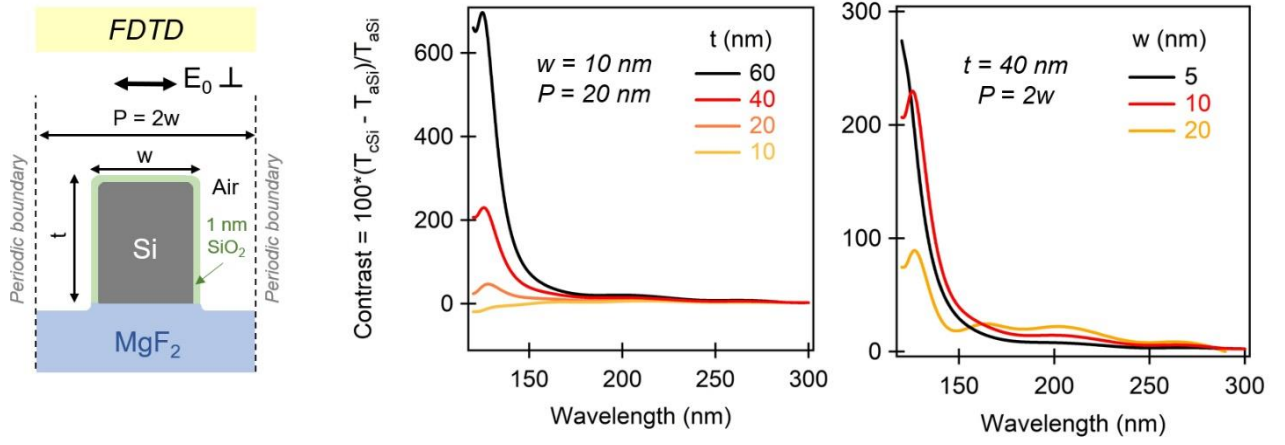

**Figure 6. ⊥ polarized a-Si to c-Si phase-change-induced transmittance contrast of the Si nanograting calculated by FDTD, as a function of its thickness t and the nanostructure width w.** Unit cell used and calculated spectra as a function of t (with w = 10 nm and P = 20 nm) and as a function of w (with t = 40 nm and P = 2w).

## 3. Conclusions and outlook

Summarizing, we have demonstrated that Si nanogratings structured at the 10-nm scale yield important VUV transmission contrasts upon a-Si to c-Si phase-change, with values of up to 600% near the wavelength of 120 nm. Such a high contrast results from the distinct interband plasmon resonances of the a-Si and c-Si nanostructures at this wavelength, where they display an epsilon-near-zero and a surface plasmonic character, respectively. These features paves the way toward UV-readable data storage platforms with a 10 to 100 times increased data density compared with single-layer CD, DVD and Blu-ray technologies. Furthermore, these platforms could be implemented by harnessing the well-established silicon nanotechnology, which readily enables the high-throughput fabrication of Si nanostructures at the 10-nm scale and below. While the presented configuration requires costly UV-transparent $MgF_2$ substrates, using other substrates or readout geometries could be considered. In this regard, effective medium modeling, which appeared to be qualitatively accurate for the Si nanograting, could enable the fast design of cost-effective device configurations.

**Methods**

The effective dielectric functions and corresponding transmittance and reflectance spectra shown in **Figure 1b** were calculated with J.A. Woollam's WVASE32 software. The dielectric functions of a-Si, c-Si were taken from [17], that of $MgF_2$ adapted from [18], and that of $SiO_2$ taken from [18], respectively. These dielectric functions are Kramers-Kronig consistent and describe dispersion effects in the spectral range considered in this work and beyond. The $MgF_2$ substrate was 1 mm thick. Therefore, values with an above $10^{-8}$ precision for the extinction coefficient k of $MgF_2$ were needed. To achieve such highly accurate values, we refined numerically the dielectric function taken from [18] so that it enables fitting the transmittance spectra of commercial 1 mm-thick MgF$_2$ substrates. The dielectric function of air was taken equal to 1. To calculate the transmittance of the Si nanograting on MgF$_2$, an approach combining coherent and incoherent transfer matrices was used, to account for coherent effects in the thin Si nanograting and multiple incoherent reflections in the MgF$_2$ susbtrate. The effect of this correction was minor. In contrast, to calculate reflectance, these reflections were discarded. Therefore the reflectance spectra shown account only for reflections by the Si nanograting. This simplification has no incidence in this work, in which the analysis focuses on transmittance data.

FDTD simulations were performed using the OPTIFDTD software, in the 2D-FDTD mode. The unit cell consisted of one period of the Si nanograting, with periodic boundary conditions being applied on its sides. Above and below the grating, the cell's dimensions were chosen so that its boundary are located markedly outside the near-field region. Perfect matching layers were placed at the top and bottom boundaries. The dielectric functions used to describe a-Si, c-Si, MgF$_2$ and air were the same as above. A 0.05 nm discretization was chosen for the mesh. The incident plane wave was launched from a plane located at the top of the cell, and two analysis planes were located at the top and the bottom of the cell to determine the transmittance and reflectance. While field maps were calculated at selected discrete wavelengths with monochromatic light, the transmittance and reflectance spectra were obtained by shining a Gaussian modulated pulse onto the structure, and calculating the Fourier transform of the transmitted and reflected pulses. The transmittance spectra were corrected to take into account multiple incoherent reflections in the 1 mm-thick MgF$_2$ substrate. The effect of this correction was minor. For the calculation of the reflectance spectra, these reflections were discarded so that the spectra shown account only for reflections by the Si nanograting. All the FDTD calculations were stable and converged.

**Acknowledgments**

This work has been partly funded by the national research grants ALPHOMENA (PID2021-123190OB-I00f) and SLIM-2P (PID2024-156974OB-C21) funded by MCIN/AEI/10.13039/501100011033 and by the European Union NextGenerationEU/PRTR and the European Regional Development Fund (ERDF).

**Supporting Information**

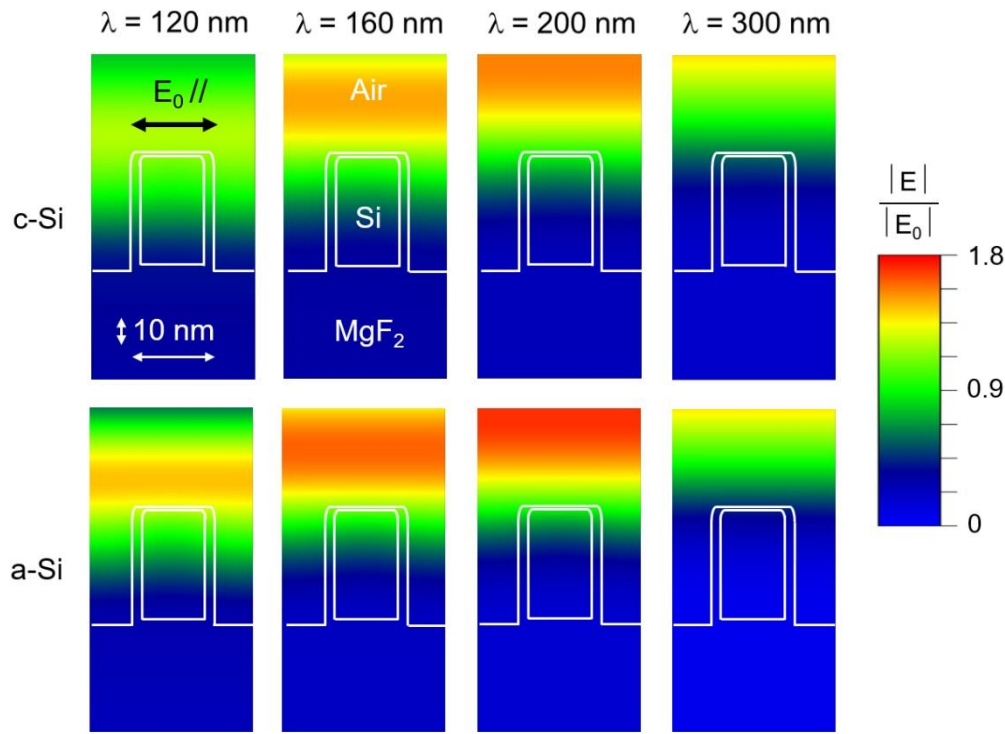

**Figure S1. // polarized electric field maps calculated by FDTD, as a function of the wavelength λ, for a c-Si and a a-Si nanograting on a MgF₂ substrate.** The unit cell and nanograting dimensions (w = 10 nm, P = 20 nm, t = 40 nm) are the same as in **Figure 2b**. For the sake of readability, the aspect ratio of the maps is not at scale. However, it is the same for all the maps to enable direct comparison between them.

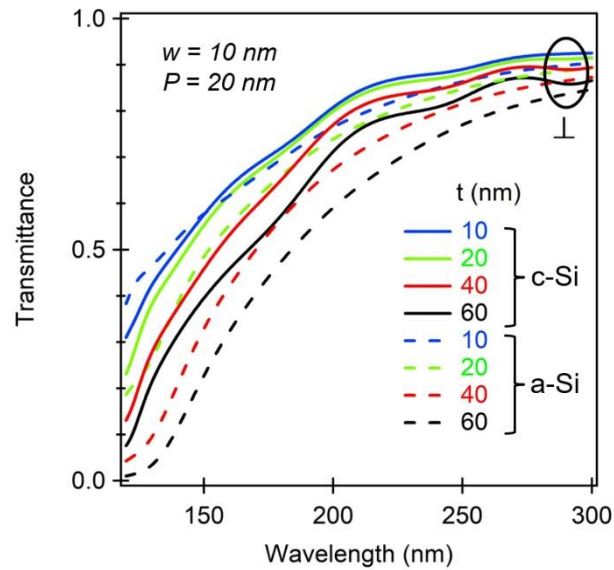

**Figure S2. ⊥ polarized transmittance spectra of the Si nanograting on a MgF₂ substrate, as calculated by FDTD as a function of its thickness t, in the a-Si and c-Si case.** The other dimensions of the nanograting were w = 10 nm and P = 20 nm.

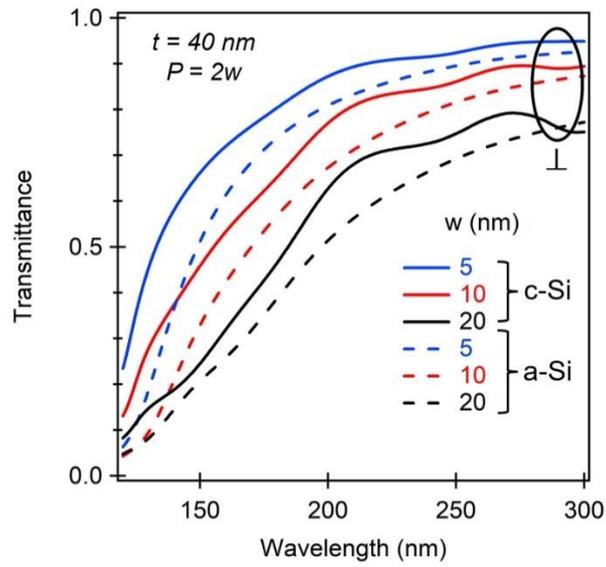

**Figure S3. ⊥ polarized transmittance spectra of the Si nanograting on a MgF$_2$ substrate, as calculated by FDTD as a function of its width w, in the a-Si and c-Si case.** The other dimensions of the nanograting were t = 40 nm and P = 2w.